\documentclass[prd,onecolumn,floatfix,superscriptaddress,showpacs,preprintnumbers,nofootinbib]{revtex4}
\usepackage{graphicx,dcolumn,bm}
\usepackage{rotating}
\usepackage{amsmath}
\usepackage{amssymb}
\usepackage{array}
\usepackage{wrapfig}
\usepackage{units}
\usepackage{color}

\newcommand{\lsim}{\buildrel < \over {_\sim}}
\newcommand{\gsim}{\buildrel > \over {_\sim}}

\newcommand{\ket}[1]{\left\lvert #1\right\rangle}
\newcommand{\bra}[1]{\left\langle #1\right\rvert}

\newcommand{\be}{\begin{equation}}
\newcommand{\ee}{\end{equation}}
\newcommand{\bee}{\begin{equation*}}
\newcommand{\eee}{\end{equation*}}
\newcommand{\bea}{\begin{eqnarray}}
\newcommand{\eea}{\end{eqnarray}}
\newcommand{\bean}{\begin{eqnarray*}}
\newcommand{\eean}{\end{eqnarray*}}

\usepackage{color}

\newcommand{\ba}{\begin{eqnarray}}
\newcommand{\ea}{\end{eqnarray}}

\newcommand{\diracslash}[1]{\not\!\! #1}

\newcommand{\nc}{\newcommand}

\nc{\newsection}[1]{\section{#1}\setcounter{equation}{0}}
\nc{\newappendix}[1]{\section*{#1}\setcounter{equation}{0}}
\nc{\scm}{\scriptscriptstyle\mathrm}
\nc{\f}{\frac}
\nc{\baa}{\begin{array}}      \nc{\eaa}{\end{array}}
\nc{\bit}{\begin{itemize}}    \nc{\eit}{\end{itemize}}
\nc{\ben}{\begin{enumerate}}  \nc{\een}{\end{enumerate}}
\nc{\bce}{\begin{center}}     \nc{\ece}{\end{center}}
\nc{\bfl}{\begin{flushright}} \nc{\efl}{\end{flushright}}
\nc{\btb}{\begin{tabular}}    \nc{\etb}{\end{tabular}}
\nc{\eps}{\varepsilon}
\nc{\vp}{\varphi}
\nc{\tvp}{\widetilde{\varphi}}
\nc{\D}{\mbox{$\not\!\!D$}}
\nc{\Db}{\mbox{${\raisebox{2mm}{\boldmath ${}^\leftarrow$}\hspace{-4mm} D}$}}
\nc{\Dfb}{\mbox{$\raisebox{2mm}{\boldmath ${}^\leftrightarrow$}\hspace{-4mm} D$}}
\nc{\vpj }{\mbox{${\vp^\dag i\,\raisebox{2mm}{\boldmath ${}^\leftrightarrow$}\hspace{-4mm} D_\mu\,\vp}$}}
\nc{\vpjt}{\mbox{${\vp^\dag i\,\raisebox{2mm}{\boldmath ${}^\leftrightarrow$}\hspace{-4mm} D_\mu^{\,I}\,\vp}$}}

\def\wt{\widetilde}
\def\gpbz{{\bar g}_\pi^{(0)}}
\def\gpbo{{\bar g}_\pi^{(1)}}
\def\gpbt{{\bar g}_\pi^{(2)}}
\def\gpbi{{\bar g}_\pi^{(i)}}

\newcommand\slurp[1]{#1}
\newcommand\sslurp[2]{#1{#2}}
{\catcode`/=\active \expandafter}%
\slurp{\newcommand/}{/\penalty1000\hskip0pt\relax}
{\catcode`:=\active \expandafter}%
\slurp{\newcommand:}{:\penalty1000\hskip0pt\relax}

\newcommand\addspace{\ifcat\nextchar a\spacefactor999. \else.\fi}
{\catcode`\.=\active \expandafter}%
\slurp{\newcommand.}{\futurelet\nextchar\addspace}

\usepackage[unicode]{hyperref} 
\ifx\href\undefined\def\href#1{}\fi
\ifx\texorpdfstring\undefined\def\texorpdfstring#1#2{#1}\fi

\newcommand\myslash{/} \newcommand\mycolon{:}
\newcommand\doi{{\catcode`/=\active \catcode`:=\active \expandafter}\sslurp\realdoi}
{\catcode`/=\active \catcode`:=\active \expandafter}%
\slurp{\newcommand\realdoi[1]{{\let/=\myslash \let:=\mycolon
                               \edef\raw{{http://dx.doi.org/#1}}\expandafter}%
                               \expandafter\href\raw{doi:#1}}}

\newcommand{\dslash}[1]{#1 \llap{/\kern-0.5pt}}
\newcommand{\Dslash}[1]{#1 \llap{/\kern+1.2pt}}
\newcommand{\DDslash}[1]{#1 \llap{/\kern+2.3pt}}
\newcommand{\dslashh}[1]{#1 \llap{/\kern+1pt}}

\begin{document}

\preprint{ACFI-T14-12}

\title{Electric Dipole Moments: A Global Analysis}



\author{Timothy Chupp}
\affiliation{{\it Physics Department, University of Michigan\\ Ann Arbor, MI 48109 USA}}
\author{Michael Ramsey-Musolf}
\affiliation{{\it Amherst Center for Fundamental Interactions\\
Department of Physics, University of Massachusetts Amherst\\
Amherst, MA 01003 USA}}
\affiliation{{\it Kellogg Radiation Laboratory, California Institute of Technology\\
Pasadena, CA 91125 USA}}

\date{\today}

\begin{abstract}
We perform a global analysis of searches for the permanent electric dipole moments (EDMs) of the neutron, neutral atoms, and molecules in terms
of six leptonic, semileptonic, and nonleptonic interactions involving photons, electrons, pions, and nucleons. Translating the results into fundamental CP-violating effective interactions through dimension six involving Standard Model particles, we obtain rough lower bounds on the scale of beyond the Standard Model CP-violating interactions ranging from 1.5 TeV for the electron EDM 
to 1300 TeV for the nuclear spin-independent electron-quark interaction. We show that future measurements involving systems or combinations of systems with complementary sensitivities to the low-energy parameters may extend the mass reach by an order of magnitude or more.

\end{abstract}

\pacs{11.30.Er, 13.40.-f, 14.60.Cd, 32.10.Dk, 33.15.Kr}
\maketitle

\section{Introduction.}
\label{sec:intro}
The search for permanent electric dipole moments (EDMs) of the neutron, atoms, and molecules provides one of the most powerful
probes of the combination of time-reversal (T) and parity (P) symmetry and the underlying combination of charge conjugation (C) and P at the elementary particle level (for recent reviews, see Refs.~\cite{Engel:2013lsa,Pospelov:2005pr}). The non-observation of the EDMs of the neutron ($d_n$) \cite{Baker:2006ts} and
$^{199}$Hg atom\cite{Griffith:2009zz} are consistent with the Standard Model CP-violation (CPV) characterized by the Cabibbo-Kobayashi-Maskawa (CKM) matrix but imply a vanishingly small coefficient ${\bar\theta}$ of the CPV $G{\widetilde G}$ operator in the SM strong interaction Lagrangian. Scenarios for physics beyond the Standard Model (BSM) typically predict the existence of new sources of CPV that -- in contrast to the CKM CPV -- do not give suppressed contributions to EDMs unless the CPV parameters themselves are small or the mass scales high. The presence of new CPV interactions is required to account for the cosmic matter-antimatter asymmetry. If the associated energy scale is not too high compared to the scale of electroweak symmetry-breaking (EWSB), and if the responsible CPV interactions are flavor diagonal, then EDMs provide a particularly important window\cite{Morrissey:2012db}.

The past decade has witnessed tremendous strides in the sensitivity of EDM searches as well as the development of prospects for even more sensitive tests. Recently, the ACME collaboration\cite{Baron:2013eja} has reported a limit on the EDM of the paramagnetic ThO molecule that 
yields an order of magnitude more stringent bounds  
on  CPV interactions than limits implied by previously reported results of YbF~\cite{Hudson:2011zz} and Tl~\cite{Regan:2002ta}. As we discuss below,
the ACME result probes BSM mass scale $\Lambda$ ranging from 1.5 TeV for the electron EDM 
to 1300 TeV for the nuclear spin-independent electron-quark interaction.
A few years earlier, a similar advance in sensitivity was achieved for $d_A(^{199}\mathrm{Hg}$) \cite{Griffith:2009zz}. Looking to the future, efforts are underway to improve the sensitivity of $d_n$ searches by one to two orders of magnitude, to achieve similar progress in neutral atoms such as Xe, Rn and Ra, and to explore the development of proton and light nuclear EDM searches using storage rings (for a recent discussion of present and future EDM search efforts, see Ref.~\cite{Kumar:2013qya}).  For $\mathcal{O}(1)$ BSM CPV phases, these experiments could probe $\Lambda$ of order $50-100$ TeV.

In this context, it is useful to try and develop a global picture of the information that has been or will be provided by present and future EDM searches. Ideally, one would like to interpret the results in terms of underlying BSM interactions in a way that would point in the direction of, or rule out, particular scenarios for new CPV. In practice, most analyses follow a more constrained approach. Theorists often work within the framework of a specific model, such as the minimal supersymmetric Standard Model, and derive constraints on the model parameters from the EDM search null results (see, {\em e.g.}, Refs.~\cite{Li:2010ax,Inoue:2014nva}). Experimental analyses, on the other hand, are often agnostic about a specific model realization but report limits on various \lq\lq sources" of an EDM ({\em e.g.}, the quark EDM or chromo-EDM; see below) assuming only one of these is present.  While entirely appropriate, such studies inherently either build in  a model-dependent bias or preclude the possibility that multiple sources may be present and, thus, may not reveal the full landscape of CPV sources probed by EDM experiments. For these reasons, it is also instructive to consider EDMs from a model-independent perspective that does not impose the \lq\lq single-source" restriction. 

In what follows, we begin this undertaking by providing a model-independent, global analysis of EDM searches. We carry out this analysis in terms of a set of low-energy hadronic and atomic parameters that one may ultimately match onto CPV interactions at the elementary particle level. It is particularly convenient to organize the latter in terms of an effective field theory (EFT) involving Standard Model degrees of freedom. The effective operators arising in the EFT constitute the CPV \lq\lq sources". In this context, the EFT provides a bridge between the atomic, nuclear, and hadronic matrix elements most directly related to the EDM searches and the possible origins of new CPV involving BSM particles and their interactions. A given BSM scenario will yield specific, model-dependent predictions for the EFT operator coefficients that one may compare with the constraints obtained from our model-independent global analysis. A detailed discussion of the EFT and its relation to both the low-energy parameters and various BSM scenarios appears in Ref.~\cite{Engel:2013lsa}, whose notation and logic we generally adopt in this paper. 

The atomic, molecular, hadronic, and nuclear matrix elements most directly related to the experimental EDMs themselves arise from a set of low-energy leptonic, semileptonic, and non-leptonic interactions. As we discuss below, the dominant contributions arise from:
$d_e$; T- and P-violating (TVPV)\footnote{The symmetry-violation studied experimentally is explicitly TVPV, rather than CPV, as the systems consists of only particles and not their antiparticles. By virtue of the CPT theorem, these observables are related to CPV interactions and the elementary particle level assuming the latter are described by a relativistic quantum field theory.} pseudoscalar-scalar and tensor electron-nucleon interactions, characterized by strengths $C_S$ and $C_T$, respectively; the isoscsalar and isovector TVPV pion-nucleon couplings ${\bar g}_\pi^{(I)}$ for $I=0,1$; and a \lq\lq short-distance" contribution to the neutron EDM, ${\bar d}_n$.  In this context, we find that:
\begin{itemize}
\item[(i)] The EDMs of paramagnetic systems are primarily sensitive to the $d_e$ and $C_S$.\footnote{This has been discussed by other authors, for example note Refs.~\cite{Pospelov:2005pr,rf:Dzuba2011,rf:JungEDMLimit}.}
\item[(ii)] Diamagnetic atom EDMs carry the strongest sensitivity to $C_T$ and the ${\bar g}_\pi^{(0,1)}$, whereas the neutron EDM depends most strongly on ${\bar d}_n$ and $\gpbz$ providing four effective CPV parameters that are constrained by results from four experimental systems. 
\item[(iii)] Inclusion of both $d_e$ and $C_S$ in the global fit yields an upper bound on each parameter that is an order of magnitude less stringent than would be obtained under the \lq\lq single-source" assumption.
\item[(iv)] Uncertainties in the nuclear theory preclude extraction of a significant limit on $\gpbo$ from $d_A(^{199}\mathrm{Hg})$, whereas the situation regarding $\gpbz$ is under better theoretical control. Including the TlF and $^{129}$Xe in the global fit leads  to an order of magnitude tighter constraint on $\gpbo$ than on $\gpbz$. 
\item[(v)] Looking to the future, a new probe of the Fr EDM with a $d_e$ sensitivity of $10^{-28}$ e-cm~\cite{rf:Gould2007} could have a significantly stronger impact on the combined $d_e$-$C_S$ global fit than would an order of magnitude improvement in the ThO sensitivity.  The addition of new, more stringent limits on the EDMs of the neutron, $^{129}$Xe atom, and $^{225}$Ra atom would lead to substantial improvements in the sensitivities to both $\gpbz$ and $\gpbo$. 
\end{itemize}
The quantitative implications of these features are summarized in Table \ref{tab:global}, where we present our results for 95\% confidence-level upper limits based on the current set of experimental results.
\begin{table}[hb]
\begin{centering}
\begin{tabular}{|l||c|}
\hline\hline
Parameter  (units)	& 95\% limit\\
\hline
$d_e$ (e-cm)		 & $5.4\times 10^{-27}$ \\ 
\hline
$C_S$			& $4.5\times 10^{-7}$ \\
\hline
$C_T$ 			 & $2\times 10^{-6}$ \\
\hline
$\bar d_n$ (e-cm) 	 & $12\times 10^{-23}$ \\
\hline
$\gpbz$ 		& $8\times 10^{-9}$ \\
\hline
$\gpbo$ 		 & $1\times 10^{-9}$ \\
\hline\hline
\end{tabular}
\caption{Ninety-five percent confidence level bounds on the six parameters characterizing the EDMs of the neutron, neutral atoms, and molecules obtained from the fit described in the text.
\label{tab:global}}
\end{centering}
\end{table}

In terms of the underlying CPV sources, it is interesting to discuss the significance of the foregoing. Among the highlights are:
\begin{itemize}
\item[(i)] The QCD vacuum angle ${\bar\theta}$ enters most strongly through $\gpbz$ and ${\bar d}_n$. From Table \ref{tab:global} and the analysis of hadronic matrix elements in Ref.~\cite{Engel:2013lsa}, we conclude that $|{\bar\theta}|\leq{\bar\theta}_\mathrm{max}$ with $2\times 10^{-7} \lsim {\bar\theta}_\mathrm{max} \lsim 1.6\times 10^{-6}$, where the bound is dominated by the constraint on $\gpbz$ and where the range is associated with the theoretical, hadronic physics uncertainty. We observe that this limit is considerably weaker than would be obtained under the \lq\lq single-source" assumption.
\item[(ii)] The quantities $d_e$ and $C_S$ are most naturally expressed in terms of $(v/\Lambda)^2$, where   $v=246$ GeV is the weak scale; the electron Yukawa coupling $Y_e$; and a set of dimensionless Wilson coefficients $\delta_e$ and $C_{eq}^{(-)}$. Since the electron EDM is a dipole operator, it carries one power of $Y_e$ whereas the semileptonic interaction does not. For a given value of the BSM scale $\Lambda$, the results in Table \ref{tab:global} implies a constraint on $C_{eq}^{(-)}$ that is roughly five hundred times more stringent than the bound on $\delta_e$. In the event that $C_{eq}^{(-)}$ and $\delta_e$ arise at tree-level and one-loop orders, respectively, the corresponding lower bound on $\Lambda$ from $C_S$ is roughly a thousand times greater than the limit extracted from $d_e$. Thus, for BSM scenarios that generate both a nonvanishing $C_{eq}^{(-)}$ and $\delta_e$, the impact of the semileptonic CPV interaction on paramagnetic atom EDMs may be considerably more pronounced than that of the electron EDM.
\item[(iii)] The bounds on $\gpbo$ are roughly ten times weaker than quoted in earlier theoretical literature, owing in part to use of a theoretically consistent computation of its contribution to the neutron EDM\cite{Seng:2014pba}. For some underlying CPV sources, such as those generated in left-right symmetric models, the dependence of diamagnetic EDMs on $\gpbo$ may be relatively more important than the dependence on $\gpbz$ due to an isospin-breaking suppression of the latter. Consequently, one may expect more relaxed constraints on CPV parameters in left-right symmetric extensions of the Standard Model (as well as scenarios that yield sizable isovector quark chromo-EDMs) than previously realized, given these less stringent bounds on $\gpbo$. 
\end{itemize} 

In the remainder of this paper, we discuss in detail the analysis leading to these conclusions. In Section \ref{sec:theory}, we summarize the theoretical framework, drawing largely on the study in Ref.~\cite{Engel:2013lsa}. Section \ref{sec:exp} summarizes the present experimental situation and future prospects. We discuss the observables and their dependence on the six parameters in Table~\ref{tab:global}. In Section \ref{sec:GlobalAnalysis} we present the details of our fitting procedure. We conclude with an outlook and discussion of the implications in Section \ref{sec:outlook}.

\section{Theoretical Framework}
\label{sec:theory}

\subsection{Low-energy parameters}
\label{sec:LEParameters}

The starting point for our analysis is the set of low-energy atomic and hadronic interactions most directly related to the EDM measurements. We distinguish two classes of systems: paramagnetic systems, namely, those having an unpaired electron spin, and diamagnetic systems, or those having no unpaired electron (including the neutron).

\vskip 0.1in

\noindent{\em Paramagnetic systems:}

\vskip 0.1in

The EDM response of paramagnetic atoms and polar molecules is dominated by the electron EDM and the nuclear spin-independent (NSID) electron-nucleon interaction. The EDM interaction for an elementary fermion is
\be
\label{eq:edmdef}
\mathcal{L}^\mathrm{EDM} = -i\sum_f\ \frac{ d_f}{2} 
{\bar f}\sigma^{\mu\nu} \gamma_5 f\ F_{\mu\nu} \ ,
\ee
where $F_{\mu\nu}$ is the electromagnetic field strength.
In the non-relativistic limit, Eq.~(\ref{eq:edmdef}) contains 
the TVPV interaction with the electric field ${\vec E}$,
\be
\label{eq:fedm}
\mathcal{L}^\mathrm{EDM}\rightarrow \sum_f {d_f}\ 
\chi^\dag_f {\vec\sigma}\chi_f \cdot {\vec E} \ ,
\ee
where $\chi_f$ is the Pauli spinor for fermion $f$ and ${\vec\sigma}$ is the 
vector of Pauli matrices.
The NSID interaction has the form 
\bea
\label{eq:NSID}
\mathcal{L}^\mathrm{NSID}_{eN} & = & -\frac{G_F}{\sqrt{2}}
{\bar e}i\gamma_5 e\ {\bar N} \left[ C_S^{(0)} +C_S^{(1)}\tau_3\right] N
\eea
where $G_F$ is the Fermi constant and $N$ is a nucleon spinor. Taking the nuclear matrix element assuming non-relativistic nucleons leads to the atomic Hamiltonian
\be
{\hat H}_S = \frac{i G_F}{\sqrt{2}}\, \delta({\vec r})\, \left [(Z+N)C_S^{(0)}+ (Z-N)C_S^{(1)}\right]\, \gamma_0\gamma_5\ \ \ ,
\ee
where a sum over all nucleons is implied and 
where the Dirac matrices act on the electron wavefunction. The resulting atomic EDM $d_A$ is then given by
\be
d_A = \rho_A^e d_e - \kappa_S^{(0)}\, C_S\ \ \ ,
\ee
where
\be
C_S\equiv C_S^{(0)} +\left( \frac{Z-N}{Z+N}\right) C_S^{(1)}
\ee
and where $\rho_A^e$ and $\kappa_S^{(0)}$ are obtained from atomic and hadronic computations. For polar molecules, the effective Hamiltonian is
\be
{\hat H}_\mathrm{mol} = \left[W_d\, d_e +W_S\,  (Z+N)C_S\right] {\vec S}\cdot{\hat n}+\cdots\ \ \ ,
\ee
where ${\vec S}$ and ${\hat n}$ denote the unpaired electron spin and unit vector along the intermolecular axis, respectively. The resulting ground state matrix element in the presence of an external electric field ${\vec E}_\mathrm{ext}$ is
\be
\bra{\mathrm{g.s.}} {\hat H}_\mathrm{mol} \ket{\mathrm{g.s.}} = \left[W_d\, d_e +W_S\, (Z+N)C_S\right] \, \eta(E_\mathrm{ext})
\ee
with 
\be
\eta(E_\mathrm{ext}) = \bra{\mathrm{g.s.}}{\vec S}\cdot{\hat n}  \ket{\mathrm{g.s.}} _{E_\mathrm{ext}}\ \ \ .
\ee
This takes into account the orientation of the internuclear axis and the internal electric field with respect to the external field, {\it i.e.} the electric polarizability of the molecule.


\vskip 0.1in

\noindent{\em Diamagnetic atoms and nucleons:}

\vskip 0.1in

The EDMs of diamagnetic atoms of present experimental interest arise from the nuclear Schiff moment, the individual nucleon EDMs,  and the nuclear-spin-dependent electron-nucleon interaction. Defining the latter as 
\bea
\label{eq:NSD}
\mathcal{L}^\mathrm{NSD}_{eN} & = & \frac{8 G_F}{\sqrt{2}}
{\bar e} \sigma_{\mu\nu} e\ v^\nu
{\bar N} \left[ C_T^{(0)} +C_T^{(1)}\tau_3\right] S^\mu N
+\cdots \ ,
\eea
where $S^\mu$ is the spin of a nucleon moving with velocity $v^\mu$ and where the $+\cdots$ indicate sub-leading contributions arising from the electron scalar $\times$ nucleon pseudoscalar interaction. The resulting Hamiltonian is
\be
{\hat H}_T = \frac{2i G_F}{\sqrt{2}}\, \delta({\vec r})\, \left [C_T^{(0)}+ C_T^{(1)}\tau_3\right]\, {\vec\sigma}_N\cdot{\vec\gamma}\ \ \ ,
\ee
where a sum over all nucleons is again implicit;  $\tau_3$ is the nucleon isospin Pauli matrix, ${\vec\sigma}_N$ is the nucleon spin Pauli matrix, and ${\vec\gamma}$ acts on the electron wave function. Including the effect of ${\hat H}_T$, the individual nucleon EDMs $d_N$, and the nuclear Schiff moment $S$, one has
\be
\label{eq:diamag}
d_A = \sum_{N=p,n} \rho_Z^N d_N + \kappa_S S - \left[k_T^{(0)} C_T^{(0)} + k_T^{(1)} C_T^{(1)}\right]\ \ \ .
\ee
A compilation of the $\rho_Z^N$, $\kappa_S$, and $k_T^{(0,1)}$ can be found in Ref.~\cite{Engel:2013lsa}\footnote{We note that the values for  the $\kappa_S$ given in that work should be multiplied by an overall factor of $-1$ given the convention used there and in Eq.~(\ref{eq:diamag}).}.

The nuclear Schiff moment arises from a TVPV nucleon-nucleon interaction generated by the pion exchange, where one of the pion-nucleon vertices is the strong pion-nucleon coupling and the other is the TVPV pion-nucleon interaction:
\be
\label{eq:pinn}
\mathcal{L}^\mathrm{TVPV}_{\pi NN} = {\bar N}\left[\gpbz{\vec\tau}\cdot{\vec\pi} +\gpbo \pi^0 + \gpbt\, \left(3\tau_3\pi^0-{\vec\tau}\cdot{\vec\pi}\right)\right] N
\ \ \ .
\ee
As discussed in detail in ~\cite{Engel:2013lsa} and references therein, the isotensor coupling $\gpbt$ is generically suppressed by a factor
$\lesssim 0.01$ with respect to $\gpbz$ and $\gpbo$ by factors associated with isospin-breaking and/or the electromagnetic interaction for underlying sources of CPV. Consequently we will omit $\gpbt$ from our analysis. The nuclear Schiff moment can then be expressed as
\be
S = \frac{m_N g_A}{F_\pi} \left[a_0\gpbz + a_1\gpbo\right]
\ee
where $g_A\approx 1.27$ is the nucleon isovector axial coupling, and $F_\pi=92.4$ MeV is the pion decay constant. The specific values of $a_{0,1}$ for the nuclei of interest are tabulated in Table~\ref{tb:SchiffCoef}. As discussed in detail in Ref.~\cite{Engel:2013lsa}, there exists considerable uncertainty in the nuclear Schiff moment calculations, so we will adopt the \lq\lq best values" and theoretical ranges for the $a_{0,1}$ given in that work. 

The neutron and proton EDMs arise from two sources. The long-range contributions from the TVPV $\pi$-$NN$ interaction have been computed using heavy baryon chiral perturbation theory, with the remaining short distance contributions contained in the \lq\lq low-energy constants" ${\bar d_n}$ and ${\bar d_p}$ \cite{Seng:2014pba}:
\bea
\label{eq:dnfull}
d_n & = & {\bar d}_n-\frac{e g_A\gpbz}{8\pi^2 F_\pi}\left\{ \ln \frac{m_\pi^2}{m_N^2} -\frac{\pi m_\pi}{2 m_N}
+\frac{\gpbo}{4\gpbz}\, (\kappa_1-\kappa_0)\frac{m_\pi^2}{m_N^2}\ln  \frac{m_\pi^2}{m_N^2}\right\}\\
\label{eq:dpfull}
d_p & = & {\bar d}_p+\frac{e g_A\gpbz}{8\pi^2 F_\pi} \left\{ \ln \frac{m_\pi^2}{m_N^2} -\frac{2\pi m_\pi}{m_N}-
\frac{\gpbo}{4\gpbz}\left[  \frac{2\pi m_\pi}{m_N}  +(\frac{5}{2}+\kappa_0+\kappa_1) \frac{m_\pi^2}{m_N^2}\ln  \frac{m_\pi^2}{m_N^2}\right]\right\} \ \ \ .
\eea
At present, we do not possess an up-to-date, consistent set of $\rho_Z^N$ for all of the diamagnetic atoms of interest here. Rather than introduce an additional set of associated nuclear theory uncertainties, we thus do not include these terms in our fit. Looking to the future, additional nuclear theory work in this regard would be advantageous since, for example, the sensitivity of the present $^{199}$Hg result to $d_n$ is not too different from the limit obtained in Ref.~\cite{Baker:2006ts}.  


\vskip 0.1in

\noindent{\em Low energy parameters: summary}

\vskip 0.1in

Based on the foregoing discussion, our global analysis of EDM searches will take into account the following parameters: 
\begin{itemize}
\item Paramagnetic atoms and polar molecules: $d_e$ and $C_S$ 
\item Neutron and diamagnetic atoms: $\gpbz$, $\gpbo$, ${\bar d}_n$, and $C_T^{(0,1)}$ for the neutron and diamagnetic atoms. 
\end{itemize}

\subsection{CPV sources of the low-energy parameters}

In order to interpret the low-energy parameters in terms of underlying sources of CPV, we will consider those contained in the SM as well as possible physics beyond the SM.  A convenient, model independent framework for doing so entails writing the CPV Lagrangian in terms of SM fields~\cite{Engel:2013lsa}: 
\be
\label{eq:LCPV1}
\mathcal{L}_\mathrm{CPV} = \mathcal{L}_\mathrm{CKM}+\mathcal{L}_{\bar\theta}
+\mathcal{L}_\mathrm{BSM}^\mathrm{eff}\ .  \ee 
Here the CPV SM CKM
\cite{Kobayashi:1973fv} and QCD
\cite{'tHooft:1976up,Jackiw:1976pf,Callan:1976je} interactions are 
\bea
\mathcal{L}_\mathrm{CKM} &=& -\frac{ig_2}{\sqrt{2}} V_\mathrm{CKM}^{pq} {\bar
U}_L^p \diracslash{W}^+ D_L^q +\mathrm{h.c.}\ , \\
\mathcal{L}_{\bar\theta} &=& -\frac{g_3^2}{16\pi^2} {\bar\theta} \,
\mathrm{Tr}\left(G^{\mu\nu}{\tilde G}_{\mu\nu}\right) \ , 
\label{thetaterm}
\eea 
where $g_2$ and $g_3$ are the weak and strong coupling constants,
respectively, $U_L^p$ ($D_L^p$) is a generation-$p$ left-handed up-type
(down-type) quark field, $V_\mathrm{CKM}^{pq}$ denotes a CKM matrix element,
$W_\mu^{\pm}$ are the charged weak gauge fields, and ${\tilde G}_{\mu\nu}
=\epsilon_{\mu\nu\alpha\beta}G^{\alpha\beta}/2$
($\epsilon_{0123}=1$ )\footnote{Note that our sign convention for
$\epsilon_{\mu\nu\alpha\beta}$, which follows that of
Ref.~\cite{Grzadkowski:2010es}, is opposite to what is used in
Ref.~\cite{Pospelov:2005pr} and elsewhere.  Consequently,
$\mathcal{L}_{\bar\theta}$ carries an overall $-1$ compared to what frequently
appears in the literature.} is the dual to the gluon field strength
$G^{\mu\nu}$.  The effects of possible BSM CPV are encoded in a tower of higher-dimension effective operators, 
\be
\label{eq:LCPV2}
\mathcal{L}_\mathrm{BSM}^\mathrm{eff}= \frac{1}{\Lambda^2}\ \sum_i
\alpha^{(6)}_i \, \mathcal{O}_i^{(6)} \, +\cdots \ , 
\ee 
where $\Lambda$ is the BSM mass scale considered to lie above the weak scale $v=246$ GeV and where
we have shown explicitly only those operators arising at dimension six.  
These operators \cite{Grzadkowski:2010es} are listed in Tables
3 and 4 of Ref.~\cite{Engel:2013lsa}. For purposes of this review, we focus on the subset listed in Table \ref{tab:CPVops}.

\begin{table}[t]
\centering \renewcommand{\arraystretch}{1.5}
\begin{tabular}{||c|c|c||}
\hline\hline
$\mathcal{O}_{\wt G}$ & $f^{ABC} \wt G_\mu^{A\nu} G_\nu^{B\rho} G_\rho^{C\mu} $ & CPV 3 gluon \\
\hline
$\mathcal{O}_{uG}$ & $(\bar Q \sigma^{\mu\nu} T^A u_R) \tvp\, G_{\mu\nu}^A$ & up-quark Chromo EDM\\
$\mathcal{O}_{dG}$ & $(\bar Q \sigma^{\mu\nu} T^A d_R) \vp\, G_{\mu\nu}^A$& down-quark Chromo EDM\\
$\mathcal{O}_{fW}$ &$(\bar F \sigma^{\mu\nu} f_R) \tau^I \Phi\, W_{\mu\nu}^I$  & fermion SU(2)$_L$ weak dipole \\
$\mathcal{O}_{fB}$ & $(\bar F \sigma^{\mu\nu} f_R) \Phi\, B_{\mu\nu}$ & fermion U(1)$_Y$ weak dipole \\
\hline
$Q_{ledq}$ & $(\bar L^j e_R)(\bar d_R Q^j)$ & CPV semi-leptonic \\
$Q_{lequ}^{(1)}$ & $(\bar L^j e_R) \epsilon_{jk} (\bar Q^k u_R)$ & \\
$Q_{lequ}^{(3)}$ & $(\bar L^j \sigma_{\mu\nu} e_R) \epsilon_{jk} (\bar Q^k
\sigma^{\mu\nu} u_R)$ &  \\
\hline
$Q_{quqd}^{(1)}$ & $(\bar Q^j u_R) \epsilon_{jk} (\bar Q^k d_R)$ & CPV four quark\\
$Q_{quqd}^{(8)}$ & $(\bar Q^j T^A u_R) \epsilon_{jk} (\bar Q^k T^A d_R)$ & \\
\hline
$Q_{\varphi ud}$ & $i\left({\tilde\varphi}^\dag D_\mu \varphi\right) {\bar u}_R\gamma^\mu d_R$ & quark-Higgs\\
\hline\hline
\end{tabular}
\caption{Dimension-six CPV operators that induce atomic, hadronic, and nuclear EDMs. Here $\varphi$ is the SM Higgs doublet, $\tvp=i\tau_2\vp^\ast$, and $\Phi=\vp$ ($\tvp$) for $I_f< 0$ ($>0$). 
\label{tab:CPVops}}
\end{table}

After EWSB, quark-gluon interactions give
rise to the quark chromo-electric dipole moment (CEDM) interaction:
\be
\label{eq:cedmdef}
\mathcal{L}^\mathrm{CEDM} = -i\sum_q\ \frac{g_3 {\tilde d}_q}{2}\ 
{\bar q} \sigma^{\mu\nu} T^A\gamma_5 q\ G_{\mu\nu}^A \ ,
\ee
where $T^A$ ($A=1, \ldots, 8$) are the generators of the color group.
Analogously, 
$Q_{f W}$ and $Q_{f B}$ generate the elementary fermion EDM interactions of Eq.~(\ref{eq:edmdef}). 
Letting 
\be
\alpha^{(6)}_{f V_k}\equiv g_k C_{f V_k} \ ,
\ee
where $V_k=$ $B$, $W$, and $G$ for $k=1,2,3$ respectively,
the relationships between the ${\tilde d}_q$ and $d_f$ and the $C_{f V_k}$ are 
\bea
\label{eq:pcdef}
{\tilde d_q} & =& - \frac{\sqrt{2}}{v} \left(\frac{ v}{\Lambda}\right)^2\ 
\mathrm{Im} \ C_{q G} \ ,
\\
\label{eq:prdef}
d_f & = &  - \frac{\sqrt{2} e}{v}\ \left(\frac{ v}{\Lambda}\right)^2\ 
\mathrm{Im}\ C_{f \gamma} \ ,
\eea
where
\be
\label{eq:Cfgammadef}
\mathrm{Im}\ C_{f \gamma} \equiv 
\mathrm{Im}\ C_{f B}+ 2I_3^f \; \mathrm{Im}\ C_{f W} \ ,
\ee
and $I_3^f$ is the third component of weak isospin for fermion $f$.  Here, we 
have expressed $d_f$ and ${\tilde d}_q$ in terms of the  Fermi scale 
$1/v$, a 
dimensionless ratio involving the BSM scales $\Lambda$ and $v$, and the dimensionless 
Wilson coefficients. Expressing these quantities in units of fm one has
\bea
{\tilde d_q} & = &  - (1.13 \times 10^{-3}\ \mathrm{fm}) 
\left(\frac{ v}{\Lambda}\right)^2\ \mathrm{Im} \ C_{q G} \ ,\\
d_f & = & - (1.13 \times 10^{-3}\ e \, \mathrm{fm})
\left(\frac{ v}{\Lambda}\right)^2\ \mathrm{Im}\ C_{f \gamma} \ .
\eea

As discussed in Ref.~\cite{Engel:2013lsa}, it
is useful to observe that the EDM and CEDM operator coefficients are 
typically proportional to the corresponding fermion masses\footnote{Exceptions to this
statement do occur.}, as the operators 
that generate them above the weak scale 
($Q_{q \wt G}$, $Q_{f \wt W}$, $Q_{f \wt B}$) contain explicit factors of the Higgs 
field dictated by electroweak gauge invariance. It is, thus, convenient to make the
dependence on the corresponding fermion Yukawa couplings $Y_f=\sqrt{2} m_f/v$
explicit 
and to define two dimensionless quantities ${\tilde\delta}_q$ and $\delta_f$ 
that embody all of the model-specific dynamics responsible for the EDM and CEDM, respectively,
apart from Yukawa insertion:
\bea
\mathrm{Im}\ C_{q G} & \equiv & Y_q\, {\tilde\delta}_q 
\rightarrow {\tilde d}_q = -(1.13 \times 10^{-3}\ \mathrm{fm})\,  
\left(\frac{v}{\Lambda}\right)^2\, Y_q\,  {\tilde\delta}_q \ ,
\label{eq:nda1prime}
\\
\mathrm{Im}\ C_{{f\gamma }} & \equiv & Y_f\, {\delta}_f 
\rightarrow d_f =  -(1.13 \times 10^{-3}\ e\, \mathrm{fm})  \, 
\left(\frac{v}{\Lambda}\right)^2\, Y_f\,  {\delta}_f \ .
\label{eq:nda1}
\eea
While one often finds bounds on the elementary fermion EDM and CEDMs quoted in 
terms of $d_f$ and ${\tilde d}_q$, the quantities $\delta_f$ and 
${\tilde\delta}_q$ are typically more appropriate when comparing with the Wilson 
coefficients of other dimension-six CPV operators (see below). One may also derive 
generic (though not air tight) expectations for the relative magnitudes of various dipole operators.
For example, for a BSM scenario that generate both quark and lepton EDMs and that does not discriminate
between them apart from the Yukawa couplings, one would expect $\delta_q\sim \delta_\ell$. On the other hand, 
the corresponding light quark EDM $d_q$ would be roughly an order of magnitude larger than that of the electron, given the 
factor of ten larger light quark Yukawa coupling\footnote{ We will neglect the light-quark mass splitting and replace
$
Y_u,\ Y_d \rightarrow Y_q \equiv \frac{\sqrt{2}{\bar m}}{v}
$
with ${\bar m}$ being the average light quark mass.}. In what follows, we will therefore quote constraints on both $d_e$ and $\delta_e$ implied
by results for paramagnetic systems; for implications of the neutron and diamagnetic results for the quark EDMs ($d_q/\delta_q$) and CEDMs (${\tilde d}_q/{\tilde\delta}_q$)
we refer the reader to Ref.~\cite{Engel:2013lsa}.

The remaining operators in Table \ref{tab:CPVops} include $\mathcal{O}_{\wt G}$, the CPV Weinberg three-gluon operator (sometimes called the gluon CEDM); a set of three semileptonic operators $Q_{ledq}$, $Q_{lequ}^{(1)}$, $Q_{lequ}^{(3)}$ ; and two four-quark operators $Q_{quqd}^{(1)}$ and $Q_{quqd}^{(8)}$. An additional four-quark CPV interaction arises from the quark-Higgs operator 
$Q_{\varphi ud} $
in Table~\ref{tab:CPVops}. After EWSB, this operator contains a $W^+_\mu {\bar u}_R\gamma^\mu d_R$ vertex that, combined with tree-level exchange of the $W$ boson, gives rise to a CPV ${\bar u}_R\gamma^\mu d_R {\bar d}_L\gamma_\mu u_R$ effective interaction. As a concrete illustration, a non-vanishing Wilson coefficient
$\mathrm{Im} C_{\varphi ud}$ naturally arises in left-right symmetric models wherein new CPV phases enter {\em via} mixing of the left- and right-handed $W$ bosons and through the rotations of the left- and right-handed quarks from the weak to mass eigenstate basis. 

The aforementioned operators will give rise to various low-energy parameters of interest to our analysis. Here we summarize a few salient features:
\begin{itemize}

\item $\mathcal{L}_\mathrm{CKM}$: At the elementary particle level, the CKM-induced quark EDMs vanish through two-loop order; the first non-zero contributions arise at three-loop order for the quarks and four-loop order for the leptons. The effects of elementary fermions in the hadronic, atomic, and molecular systems of interest here are, thus, highly suppressed. The dominant contribution enters the neutron EDM and nuclear Schiff moments {\em via} the induced CPV penguin operators that generate TVPV strangeness changing meson-nucleon couplings. The expected magnitudes of $d_n$ and diamagnetic-atom EDMs are well below the expected sensitivities of future experiments, so we will not consider the effects of $\mathcal{L}_\mathrm{CKM}$ further here. 

\item $\mathcal{L}_{\bar\theta}$: The QCD $\theta$-term will directly induce a nucleon EDM as well as the TVPV coupling $\gpbz$ at leading order. Since one may rotate away $\bar\theta$ when either of the light quark masses vanish, the contributions of ${\bar\theta}$ to $d_n$ and $\gpbz$ are proportional to the square of the pion mass $m_\pi^2$. Chiral symmetry considerations imply that the effect on $\gpbo$ and $\gpbt$ is suppressed by an additional power of $m_\pi^2$ while $\gpbt$ is further reduced by the presence of isospin breaking. 

\item $\mathcal{L}_\mathrm{BSM}^\mathrm{eff}$: The presence of the quark CEDM, three-gluon operator, and CPV four-quark operators will induce non-vanishing nucleon EDMs.
As noted in section~\ref{sec:LEParameters}  the expected magnitude of $\gpbt$ relative to $\gpbz$ and $\gpbo$ is always suppressed by a factor
$\lesssim 0.01$ associated with isospin breaking and only the CPV $\pi$-$NN$ coupling constants $\gpbz$ and $\gpbo$ are included in our analysis. Additionally, the effect of a non-vanishing $\mathrm{Im} C_{\varphi ud}$ will generate both a nucleon EDM and, to leading order in chiral counting, contribute to $\gpbo$. As indicated by Eq.~(\ref{eq:dnfull}) the long-range contribution to $d_n$ associated with $\gpbo$ is suppressed by $m_\pi^2/m_N^2$, whereas the effect on $d_p$ appears at one order lower in $m_\pi/m_N$. For the diamagnetic atoms, the nuclear theory uncertainties associated with the $\gpbo$ contribution to the $^{199}$Hg Schiff moment are particularly large. At present, the sign of $a_1$ is undetermined, and it is possible that 
its magnitude may be vanishingly small\cite{Engel:2013lsa}. In contrast, the computations of $a_1$ for other diamagnetic systems,  appear to be on firmer ground.

\item The semi-leptonic operators $\mathcal{O}_{\ell e dq}$ and $\mathcal{O}_{\ell equ}^{(1,3)}$ will induce an effective nucleon spin-independent (NSID) electron-nucleon interaction
 The coefficients $C_S^{(0,1)}$ can be expressed in terms of the underlying semileptonic operator coefficients and the nucleon scalar form factors:
\begin{eqnarray}
C_S^{(0)}  &=& -g_S^{(0)}\, \left(\frac{v}{\Lambda}\right)^2\,  
\mathrm{Im}\ C_{eq}^{(-)}\\
\nonumber
C_S^{(1)}  &=&  g_S^{(1)}\, \left(\frac{v}{\Lambda}\right)^2\,  
\mathrm{Im}\ C_{eq}^{(+)}  \\
\nonumber
C_T^{(0)} & = & -g_T^{(0)}\, \left(\frac{v}{\Lambda}\right)^2\,  
\mathrm{Im}\ C_{\ell e qu}^{(3)}\\
\nonumber
C_T^{(1)} & = & -g_T^{(1)}\, \left(\frac{v}{\Lambda}\right)^2\,  
\mathrm{Im}\ C_{\ell e qu}^{(3)}
\label{eq:CSi}
\end{eqnarray}
where 
\be
\label{eq:Ceqdef}
C_{eq}^{(\pm)}= C_{\ell e dq} \pm C_{\ell e q u}^{(1)} \ \ \ .
\ee 
and the isoscalar and isovector  form factors $g_\Gamma^{(0,1)}$ are given by
\bea
\label{eq:ffdef}
\frac{1}{2} \bra{N} \left[{\bar u} \Gamma u + {\bar d}\Gamma d\right]\ket{N} 
&\equiv& g_\Gamma^{(0)} {\bar \psi_N} \Gamma \psi_N\ ,\\
\frac{1}{2} \bra{N} \left[{\bar u} \Gamma u - {\bar d}\Gamma d\right]\ket{N} 
&\equiv& g_\Gamma^{(1)} {\bar \psi_N} \Gamma \tau_3 \psi_N\ ,
\eea
where $\Gamma = 1$ and $\sigma_{\mu\nu}$, respectively. Values for these form factors can be obtained from Ref.~\cite{Engel:2013lsa}.

\item We observe that there exist more CPV sources than independent low-energy observables. Restricting one's attention to interactions of mass dimension six or less involving only the first generation fermions and massless gauge bosons, one finds thirteen independent operators. For the paramagnetic systems, the situation is somewhat simplified, as there exist only three relevant operators: the electron EDM and the two scalar (quark) $\times $ pseudscalar (electron) interactions. For the systems of experimental interest, the electron EDM and $C_S^{(0)}$ operators dominate. For the diamagnetic systems, on the other hand, there exist ten underlying CPV sources that may give rise to the quantities  $\gpbz$, $\gpbo$, ${\bar d}_n$, and $C_T^{(0,1)}$. Even with the possible addition of a future proton EDM constraint, thereby adding one additional low-energy parameter ${\bar d}_p$, it would not be possible to disentangle all ten sources from the experimentally accessible quantities. Future searches for the EDMs of light nuclei may provide additional handles (see, {\em e.g.} Ref.~\cite{Engel:2013lsa} and references therein), but an analysis of the prospects goes beyond the scope of the present study. Instead, we concentrate on the present and prospective constraints on the dominant low-energy parameters
$d_e$, $C_S$, $\gpbz$, $\gpbo$, ${\bar d}_n$, and $C_T^{(0,1)}$. 

As indicated above, the two tensor couplings depend on the same Wilson coefficient $\mathrm{Im}\ C_{\ell e qu}^{(3)}$, so their values differ only through the values of the nucleon form factors $g_T^{(0,1)}$. At this level of interpretation, a meaningful fit would include only one parameter rather than two distinct and independent tensor couplings. Unfortunately, we presently possess limited information on the nucleon tensor form factors $g_T^{(0,1)}$, and the theoretical uncertainties associated with existing computations are sizable. Consequently, we adopt an interim strategy until refined computations of the tensor form factors are available, retaining only $C_T^{(0)}$ in the fit. Henceforth, we denote this parameter by \lq\lq $C_T$".

\end{itemize}

\section{Experimental Status and Prospects}
\label{sec:exp}

Over the past six decades, a large number of EDM measurements in a variety of systems have provided results, all of which are consistent with zero. The most recent or best result for each system used in our analysis is presented in Table~\ref{tb:Results}.
The results are separated into two distinct categories as indicated above: (a) paramagnetic atoms and molecules and (b) diamagnetic systems (including the neutron). 
Although paramagnetic systems (Cs, Tl, YbF and ThO) are most sensitive to both the electron EDM $d_e$ and the nuclear spin-independent component of the electron-nucleus coupling ($C_S$), most experimenters have presented their results as a measurement of $d_e$, which requires the  assumption that $C_S=0$. As we discuss below, this assumption is not required in a global analysis of EDM results. 

Diamagnetic systems, including $^{129}$Xe and $^{199}$Hg atoms, the molecule TlF, and the neutron, are most sensitive to purely hadronic CPV sources, as well as the tensor component of the electron-nucleus coupling $C_T$ for atoms and molecules; however the electron EDM and $C_S$ contribute to the diamagnetic atoms in higher order. 
The constraints provided by the diamagnetic systems are expected to change significantly within the next few years.  Strong efforts or proposals at several labs foresee improving the neutron-EDM sensitivity by one or more orders of magnitude~\cite{rf:neutronproposalILL,rf:neutronproposalPSI,rf:neutronproposalSNS,rf:neutronproposalTRIUMF,rf:neutronproposalMunich,rf:neutronproposalSerebrov}, 
 and the EDM of $^{129}$Xe by several orders of magnitude~\cite{rf:XenonRomalis,rf:XenonTUM}. Most importantly, there has been significant progress in theory and towards a measurement of the EDMs of heavy atoms with octupole-deformed nuclei, {\it i.e.} in $^{225}$Ra~\cite{rf:RadiumRef} and $^{221}$Rn or $^{223}$Rn\cite{rf:RadonRef}. In these systems, the nuclear structure effects are expected to enhance the Schiff moment generated by the long-range TVPV pion-exchange interaction, leading to an atomic EDM 2-3 orders of magnitude larger than $^{199}$Hg. As we show below, an atomic-EDM measurement at the 10$^{-26}$ e-cm level will provide additional input that will significantly impact our knowledge of the TVPV hadronic parameters.

\subsection{Constraints on TVPV Couplings}
\label{sec:GlobalAnalysis}

From the arguments presented above, there are seven dominant effective-field-theory parameters: $d_e$, $C_S$, $C_T$, $\gpbz$, $\gpbo$, and the two isospin components of the short-range hadronic contributions to the neutron and proton EDMs, which we isolate as $\bar d_n$ and $\bar d_p$ in eq.~\ref{eq:dpfull}. We, thus, write the the EDM of a particular system as
\begin{equation}
d=\alpha_{d_e} d_e + \alpha_{C_S} C_S + \alpha_{C_T} C_T + \alpha_{\bar d_n} \bar d_n + \alpha_{\bar d_p} \bar d_p  + \alpha_{g_\pi^0} \bar g_\pi^0 + \alpha_{g_\pi^1}  \bar g_\pi^1 ,
\end{equation}
where $\alpha_{d_e}=\partial d/\partial d_e$, {\it etc.}. This can be compactly written as
\begin{equation}
d_i=\sum_{j} \alpha_{ij} C_j,
\label{eq:d_i}
\end{equation}
where $i$ labels the system, and $j$ labels the physical contribution. 
The coefficients $\alpha_{ij}$ are provided by atomic and nuclear theory calculations and are listed in Tables~\ref{tb:paramagnetics} and~\ref{tb:diamagnetics} for diamagnetic and paramagnetic systems, respectively.
The sensitivity of the EDM for each experimental system to the parameters presented as a best value and a reasonable range as set forth in Ref.~\cite{Engel:2013lsa}. 
 
\begin{table}
\centering
\begin{tabular}{|c|c|lc|}
\hline\hline
System & Year/ref &\multicolumn{2}{c|}{Result}  \\
\hline
\multicolumn{4}{|c|}{Paramagnetic systemss}\\
\hline
Cs  & 1989~\cite{Murthy:1989zz} & $d_A=(-1.8\pm6.9)\times 10^{-24}$ & e-cm \\
&& $d_e=(-1.5\pm 5.6)\times 10^{-26}$ & e-cm    \\
\hline
Tl  & 2002~\cite{Regan:2002ta}  &$d_A=(-4.0\pm 4.3)\times 10^{-25}$ & e-cm \\
& &  $d_e=(\quad 6.9\pm 7.4)\times 10^{-28}$ &  e-cm  \\
 \hline
YbF & 2011~\cite{Hudson:2011zz}  &  $d_e=(-2.4\pm 5.9)\times 10^{-28}$  &  e-cm \\
&&&\\
\hline
ThO & 2014~\cite{Baron:2013eja} &  $\omega^{\mathcal NE}=2.6\pm 5.8$ & mrad/s     \\
  &   &  $d_e=(-2.1\pm 4.5)\times 10^{-29}$ & e-cm    \\
  &   &  $C_S=(-1.3\pm 3.0)\times 10^{-9}$ &   \\  
\hline
\multicolumn{4}{|c|}{Diamagnetic systems}\\
\hline 
$^{199}$Hg & 2006~\cite{Griffith:2009zz} & $d_A=(0.49\pm 1.5)\times 10^{-29}$ & e-cm \\
\hline
$^{129}$Xe &2001~\cite{rf:Rosenberry} &  $d_A=(0.7\pm 3)\times 10^{-27}$ & e-cm\\
\hline
TlF & 2000~\cite{rf:Cho} & $d_{\rm\ }=(-1.7\pm 2.9)\times 10^{-23}$ & e-cm \\
\hline
neutron & 2006~\cite{Baker:2006ts} & $d_n=(0.2\pm1.7)\times 10^{-26}$   & e-cm\\

\hline
\hline 
\end{tabular}
\caption{EDM results used in our analysis as presented by the authors. When $d_e$ is presented, the assumption is $C_S=0$, and for ThO, the $C_S$ result assumes $d_e=0$. We have combined statistical and systematic errors in quadrature for cases where they are separately reported by the experimenters.   \label{tb:Results}}
\end{table}

\subsection{Paramagnetic systems: limits on $d_e$ and $C_S$}

\begin{figure}[h]
\centerline{\includegraphics[width=6in]{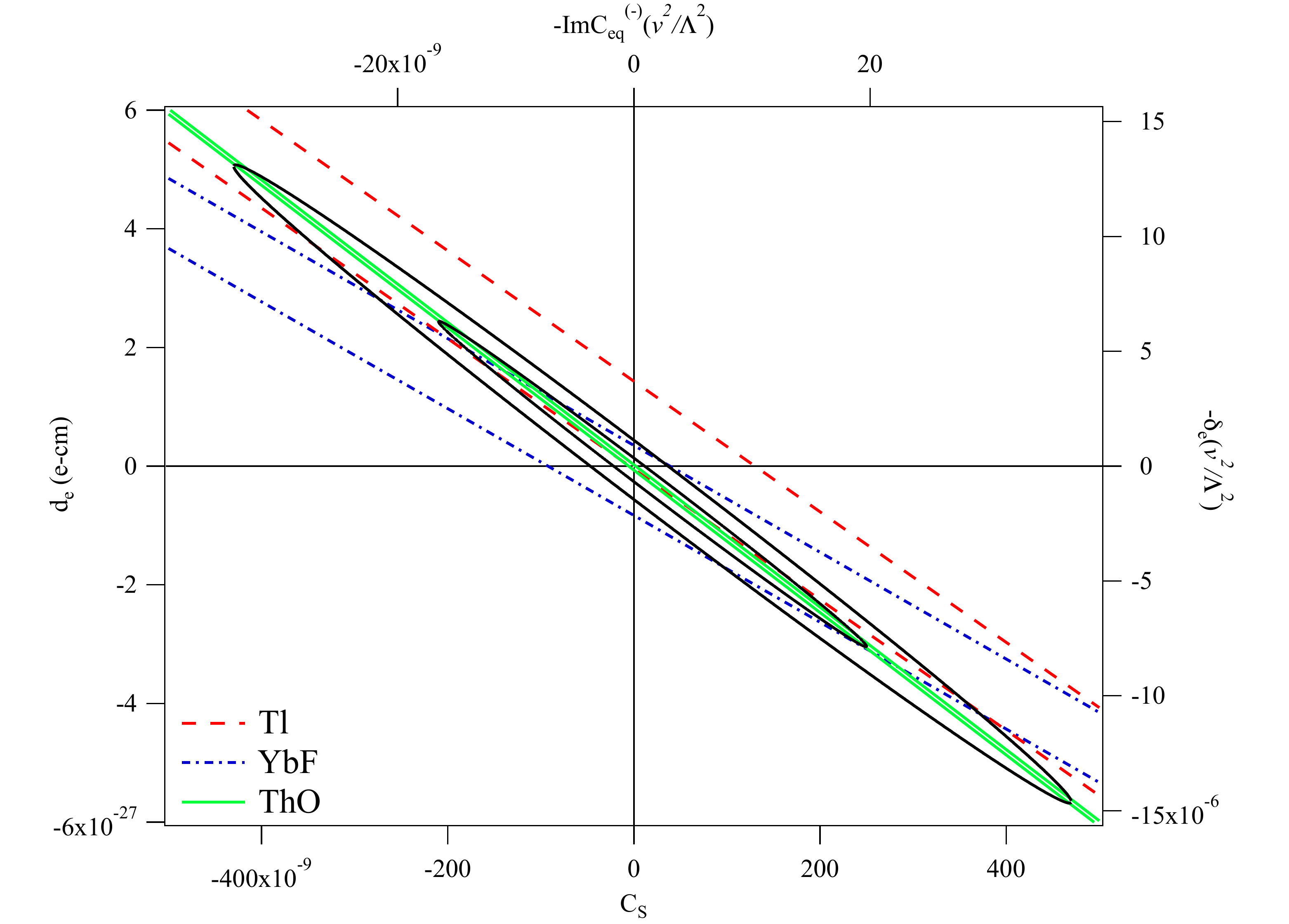}}
\caption{Electron edm $d_e$ as a function of $C_S$ from the experimental results in Tl, YbF and ThO. Also shown are 68\% and 95\% error ellipses representing the best-fit for the paramagnetic systems and including $d_A(^{199}$Hg) as discussed in the text. Also shown are the constraints on the dimensionless Wilson coefficients $\delta_e$ and $\mathrm{Im}\, C_{eq}^{(-)}$ times the squared scale ratio $(v/\Lambda)^2$.}
\label{fig:Paramagnetics}
\end{figure}

Paramagnetic systems are dominantly sensitive to $d_e$ and $C_S$; thus for Cs, Tl, YbF and ThO, following Ref.~\cite{rf:Dzuba2011} and recalling that the experimental result is reported as a limit on the electron EDM, we can define an effective electron EDM entering paramagnetic systems as
 \begin{equation}
 d^\mathrm{eff}_\mathrm{para}\approx d_e + {\alpha_{C_S}\over \alpha_{d_e}} C_S.
 \label{eq:deEquation}
 \end{equation} 
The quantities $\alpha_{C_S}/ \alpha_{d_e}$ listed in Table~\ref{tb:paramagnetics} vary over a small range, {\it i.e.} from $(0.6-1.5)\times 10^{-20}$ e-cm for the paramagnetic systems and from $(3-5)\times 10^{-20}$ for Hg, Xe and TlF. We note, as pointed out in Ref.~\cite{rf:Dzuba2011}, that while there is a significant range of $\alpha_{d_e}$ and $\alpha_{C_S}$ from different authors, there is much less dispersion in the ratio ${\alpha_{C_S}/\alpha_{d_e}}$ as reflected  in Table~\ref{tb:paramagnetics}. In  Figure~\ref{fig:Paramagnetics}, we plot $d_e$ as a function of $C_S$ using experimental results for $d^{exp}_\mathrm{para}$ for Tl, YbF and ThO. 

Constraints on $d_e$ and $C_S$ are found from a fit to the form Eq.~(\ref{eq:deEquation}) for the four paramagnetic systems listed in Table~\ref{tb:Results}. The results are \begin{equation}
d_e=(-0.4\pm 2.2)\times 10^{-27}\ {\rm e-cm}
\quad
C_S=(0.3\pm 1.7)\times 10^{-7}\quad  {\rm Best\ coefficient \ values}. 
\nonumber
\end{equation}
In order to account for the variation of atomic theory results we vary $\alpha_{C_S}/\alpha_{d_e}$ over the ranges presented in Table~\ref{tb:paramagnetics} and find that when the $\alpha_{C_S}/\alpha_{d_e}$ are most similar,
\begin{equation}
d_e=(-0.3\pm 3.0)\times 10^{-27}\ {\rm e-cm}
\quad
C_S=(0.2\pm 2.5)\times 10^{-7}\quad  {\rm Varied\ coefficient\ values}. 
\nonumber
\end{equation}

\begin{table}
\centering
\begin{tabular}{|c|c|c|c|}
\hline\hline
System & $\alpha_{d_e}$  & $\alpha_{C_S}$ & $\alpha_{C_S}/\alpha_{d_e}$ (e-cm)\\
\hline
Cs  & 123  & $7.1\times 10^{-19}$ e-cm &  $5.8\times 10^{-21}$     \\
&$(100-138)$ &  $(7.0-7.2)$  &  $(0.6-0.7)\times 10^{-20}$     \\
\hline
Tl  & -573 &  $-7\times 10^{-18}$ e-cm &  $1.2\times 10^{-20}$  \\
& $-(562-716)$ &  $-(5-9)$      &  $(1.1-1.2)\times 10^{-20}$   \\
 \hline
YbF  & $-1.1\times 10^{25}$ Hz/e-cm & $-9.2\times 10^4$ Hz &  $8.6\times 10^{-21}$   \\
 &-(0.9-1.2)& -(92-132)    &  $(8.0-9.0)\times 10^{-21}$     \\
\hline
ThO &$-5.0\times 10^{25}$ Hz/e-cm &$-6.6\times 10^5$ Hz&  $1.3\times 10^{-20}$     \\
 & $-(4.0-5.0)$ & -(4.6-6.6)  &  $(1.2-1.3)\times 10^{-20}$     \\
\hline\hline
\end{tabular}
\caption{Sensitivity  to $d_e$ and $C_S$  and the ratio $\alpha_{C_S}/\alpha_{d_e}$  for observables in paramagnetic systems  based on atomic theory calculations.  Ranges (bottom entry) for coefficients $\alpha_{ij}$ representing the contribution of each of the TVPV parameters to the observed EDM of each system. See Refs.~\cite{Ginges:2003qt,Engel:2013lsa} for  Cs and Tl. For YbF,  theory results are compiled in Ref.~\cite{rf:Dzuba2011}, and for ThO we use result from Refs.~\cite{Meyer:2008gc,rf:Dzuba2011,rf:Skripnikov2013}.  \label{tb:paramagnetics}}
\end{table}

It is in principle possible to include the diamagnetic systems, in particular $^{199}$Hg,  in constraining $d_e$ and $C_S$. To do so, however,  requires accounting for the hadronic and $C_T$ contributions to $d_A(^{199}{\rm Hg})$. As described below, the hadronic parameters and $C_T$ are constrained by our analysis of the diamagnetic systems, though the constraints are quite weak due to the limitations of both experimental input and hadronic theory. Using the experimental result for $d_A(^{199}{\rm Hg})$ combined with the upper limits for $C_T$, $\gpbz$ and $\gpbo$, we estimate the contribution to $d_A(^{199}{\rm Hg})$ from $d_e$ and $C_S$, {\it i.e.}
\begin{equation}
\alpha_{d_e}d_e+\alpha_{d_e}d_e=d_A(^{199}{\rm Hg})-(\alpha_{C_T}C_T+\alpha_{\gpbz}\gpbz+\alpha_{\gpbo}\gpbo)\approx (1.2\pm 8.0)\times 10^{-26}\ {\rm e-cm},
\end{equation}
where the coefficients $\alpha_{ij}$ for $^{199}$Hg are given in Table~\ref{tb:diamagnetics}. The large numerical value follows from the uncertainties on the parameters $C_T$, $\gpbz$ and $\gpbo$ resulting from the global fit. When this additional constraint is included, the limits on $d_e$ and $C_S$ improve slightly due to the lever arm provided by the significantly different ${\alpha_{C_S}/ \alpha_{d_e}}$ compared to the paramagnetic systems with the result
\begin{equation}
\nonumber
d_e=(-0.3\pm 2.7)\times 10^{-27}\ {\rm e-cm}
\quad
C_S=(0.2\pm 2.3)\times 10^{-7}\quad {\rm including\ ^{199}Hg}. 
\end{equation}
The 68\% and 95\% upper limits for the are
\begin{equation}
\nonumber
|d_e|=<(2.7/5.4)\times 10^{-27}\ {\rm e-cm}
\quad 
|C_S|<(2.3/4.5)\times 10^{-7}\quad (68\%/95\%)\ {\rm CL}
\nonumber
\end{equation}
Error ellipses representing 68\% and 95\% confidence interval for the two parameters $d_e$ and $C_S$ 
are presented in Figure~\ref{fig:Paramagnetics}.
The corresponding constraints on $\delta_e(v/\Lambda)^2$ and $\mathrm{Im}\, C_{eq}^{(-)} (v/\Lambda)^2$  are obtained from those for $d_e$ and $C_S$ by dividing by $-3.2\times 10^{-22}$ e-cm and $-12.7$, respectively.


\subsection{Hadronic parameters and $C_T$}

Diamagnetic atom EDMs are most sensitive to the hadronic parameters $\gpbz$ and $\gpbo$ and the electron-nucleon contribution $C_T$. As noted above, $d_e$ and $C_S$ contribute to diamagnetic systems in higher order.  Given that $d_e$ and $C_S$ are effectively constrained by the paramagnetic systems, constraints on the four free parameters $C_T$, $\gpbz$, $\gpbo$ and $\bar d_n$ are provided by four experimental results from  TlF, $^{129}$Xe and $^{199}$Hg and the neutron.  For example, the solution using the experimental centroids and the best values for the coefficients are labeled as ``exact solution'' in the first line of Table~\ref{tab:gpiCTdn}. In order to provide estimates of the constrained ranges of the parameters, we define $\chi^2$
for a given set of coefficients $\alpha_{ij}$ and a set of parameters ${\bf C_j}$:
 \begin{equation}
 \chi^2({\bf C_j})=\sum_i {(d_i^\mathrm{exp}-d_i)^2\over \sigma_{d_i^\mathrm{exp}}^2},
 \end{equation}
 where $d_i$ is given in equation~\ref{eq:d_i}. 
 We then take the following steps:
\begin{enumerate}
\item Fix $d_e$ and $C_S$ using paramagnetic systems only: $d_e=(-0.1\pm 1.8)\times 10^{-27}$ e-cm; $C_S=(0.1\pm 1.3) 10^{-7}$.
\item Vary ${\bf C_j}$ to determine $\chi^2$ contours for a specific set of $\alpha_{ij}$. For 68\% confidence and four parameters, $(\chi^2-\chi^2_{min})<4.7$. (Note that $\chi^2_{min}=0$.)
\item This procedure is repeated for values of $\alpha_{ij}$ spanning the reasonable ranges presented in Table~\ref{tb:diamagnetics} to estimate ranges  $C_T$, $\gpbz$,  $\gpbo$, and $\bar d_n$.
\end{enumerate}
Our estimates of the constraints are presented as ranges in Table~\ref{tab:gpiCTdn}.
Finally, we use the ranges for $C_T$, $\gpbz$ and $\gpbo$ to determine their contribution to the EDM of $^{199}$Hg and subtract to isolate the $d_e$/$C_S$ contribution as described above.


 \begin{table}[t]
\centering \renewcommand{\arraystretch}{1.5}
\begin{tabular}{||c|c|c|c|c|c|c|c||}
\hline\hline
System & $\partial d^{exp}/ \partial d_e$  & $\partial d^{exp}/\partial C_S$  & $\partial d^{exp}/\partial C_T$  & $\partial d^{exp}/\partial g_\pi^0$  & $\partial d^{exp}/\partial g_\pi^1$ \\
\hline $^{199}$Hg & -0.014  & $-5.9\times 10^{-22}$  & $-2\times 10^{-20}$   & $-3.8\times 10^{-18}$  & 0 \\
 & &   -0.014 - (-0.012)   & $(-5.9-(-2.0))\times 10^{-20}$   &   $(-27-(-1.9))\times 10^{-18}$  & $(-4.9 - 1.6)\times 10^{-17}$   \\
\hline
$^{129}$Xe  & -0.0008  & $-4.4\times 10^{-23}$  & $4\times 10^{-21}$   & $-2.9\times 10^{-19}$  & $-2.2\times 10^{-19}$  \\
 &    &                               & $(4-6)\times 10^{-21}$   &    $(-26-(-1.8))\times 10^{-19}$  & $(-19- (-1.1))\times 10^{-19}$   \\
\hline
TlF  & 81  &   $2.9\times 10^{-18}$  & $1.1\times 10^{-16}$  & $1.2\times 10^{-14}$  & $-1.6\times 10^{-13}$ \\
 &    &                                 & & &  \\
\hline
neutron    &  & & &  $1.5\times 10^{-14}$  & $1.4\times 10^{-16}$  \\
 &    &                                 & & &  \\
 \hline\hline
\end{tabular}
\caption{Coefficients for P-odd/T-odd parameter contributions to EDMs  for diamagnetic systems.   The $\gpbz$ and $\gpbo$ coefficients are based on data provided in Table~\ref{tb:SchiffCoef}.
\label{tb:diamagnetics}}
\end{table}

\begin{table}[hb]
\begin{center}
\begin{tabular}{|c|c|c|c|c|}
\hline\hline
System & $\kappa_S=\frac{d}{S}$ (cm/fm$^3$)  & $a_0=\frac{S}{13.5\bar g_\pi^0}$ ($e$-fm$^3$) & $a_1=\frac{S}{13.5\bar g_\pi^1}$ ($e$-fm$^3$) & $a_2=\frac{S}{13.5\bar g_\pi^2}$ ($e$-fm$^3$)  \\
\hline
TlF &  $-7.4\times 10^{-14}$~\cite{rf:TlFTheory} & -0.0124 & 0.1612 & -0.0248\\
\hline
Hg &   $-2.8/-4.0\times 10^{-17}$~\cite{rf:DzubaPRAv60p02111a2002,rf:FlbmKhrNuclPhysAp449a1986} & 0.01 (0.005-0.05) & $\pm$0.02 (-0.03-0.09) & 0.02 (0.01-0.06)\\
\hline
Xe &	   $0.27/0.38\times 10^{-17}$~\cite{rf:DzubaPLBv154p93a1985,rf:DzubaPRAv60p02111a2002} & -0.008 (-0.005-(-0.05)) & -0.006 (-0.003-(-0.05)) & -0.009 (-0.005-(-0.1))\\
\hline
Ra &   $-8.5 (-7/-8.5)\times 10^{-17}$~\cite{rf:SpevakPRCv56p1357a1997,rf:DzubaPRAv60p02111a2002} & -1.5 (-6-(-1)) & +6.0  (4-24) & -4.0 (-15-(-3)) \\
\hline\hline
\end{tabular}
\end{center}
\caption{\label{tb:SchiffCoef} Best values and ranges (in parenthesis) for atomic EDM sensitivity to the Schiff-moment and dependence of the Schiff moments on $\gpbz$ and $\gpbo$ as presented  in Ref.~\cite{Engel:2013lsa}.}
\end{table}

\begin{table}
\begin{tabular}{||c|c|c|c|c||}
\hline\hline
&&&&\\
& $C_T\times 10^7$ &$\gpbz$ & $\gpbo$& $\bar d_n$ (e-cm) \\
\hline
Exact solution & $1.265$ &$-6.687\times 10^{-10}$ & $1.4308\times 10^{-10}$& $9.878\times 10^{-24}$ \\
\hline
Range from best values of $\alpha_{ij}$ & $(-7.6-9.5)$ &$(-5.0-4.0)\times 10^{-9}$ & $(-0.2-0.4)\times 10^{-9}$& $(-5.9-7.4)\times 10^{-23}$\\
\hline
Range from best values &&&&\\
with $\alpha_{g_\pi^1}(\mathrm{Hg})=-4.9\times 10^{-17}$  & $(-7.6-8.4)$ &$(-7.0-4.0)\times 10^{-9}$ & $(0-0.2)\times 10^{-9}$& $(5.9-10.4)\times 10^{-23}$\\
\hline
Range from best values &&&&\\
with $\alpha_{g_\pi^1}(\mathrm{Hg})=+1.6\times 10^{-17}$  & $(-9.2-12.4)$ &$(-4.0-4.0)\times 10^{-9}$ & $(-0.4-0.8)\times 10^{-9}$& $(-5.9-5.9)\times 10^{-23}$ \\
\hline
Range from full variation of $\alpha_{ij}$  & $(-10.8-15.6)$ &$(-10.0-8.1)\times 10^{-9}$ & $(-0.6-1.2)\times 10^{-9}$& $(-12.0-14.8)\times 10^{-23}$\\

\hline\hline
\end{tabular}
\caption{\label{tab:gpiCTdn}Values and ranges for coefficients for diamagnetic systems and the neutron. The first line is the exact solution using the central value for each of the four experimental results; the second line is the 68\% CL range allowed by experiment combined with the best values of the coefficients $\alpha_{ij}$; the last three lines provide an estimate of the constraints accounting for the variations of the $\alpha_{ij}$ within reasonable ranges of the coefficients $\alpha_{ij}$ \cite{Engel:2013lsa}.}
\end{table}

\section{Experimental Outlook \& Theoretical Implications}
\label{sec:outlook}



Anticipated advances of both theory and experiment would lead to much tighter constraints on the TVPV parameters. The disparity shown in Table~\ref{tab:gpiCTdn} between the ranges provided by the best values of the coefficients $\alpha_{ij}$ and those provided by allowing the coefficients to vary over the reasonable ranges emphasizes the importance of improving the nuclear physics calculations, particularly the Schiff moment calculations for  $^{199}$Hg.

On the experimental front, we anticipate the following:
\begin{enumerate}
\item Increased sensitivity of the paramagnetic ThO experiment~\cite{Baron:2013eja} 
\item Improvement of up to two orders of magnitude for the the neutron-EDM~\cite{rf:neutronproposalILL,rf:neutronproposalPSI,rf:neutronproposalSNS,rf:neutronproposalTRIUMF,rf:neutronproposalMunich,rf:neutronproposalSerebrov}
\item 2-3 orders of magnitude improvement for  $^{129}$Xe\cite{rf:XenonRomalis,rf:XenonTUM,rf:MunichSheildedRoomRSI}
\item New diamagnetic atom EDM measurements from the octupole enhanced systems $^{225}$Ra~\cite{rf:RadiumRef} and $^{221}$Rn/$^{223}$Rn\cite{rf:RadonRef}
\item Possible new paramagnetic atom EDM measurement from Fr~\cite{rf:Gould2007} and Cs~\cite{rf:DavidWeissCsEDM}
\item Plans to develop storage-ring experiments to measure the EDMs of the proton and light nuclei $^2$H and $^3$He~\cite{rf:StorageRingEDMRef}
\end{enumerate}

Some scenarios for improved experimental sensitivity and their impact are presented in Table~\ref{tab:Outlook}. In the first line we summarize the current upper limits on the parameters at the 95\% CL.  The remainder of the table lists the impact of one or more experiments with the improved sensitivity noted in the third column, assuming a central value of zero. Note that we do not consider a possible future proton EDM search.
While every experiment has the potential for discovery in the sense that improving any current limit takes one into new territory, it is clear from Table~\ref{tab:Outlook} that inclusions of new systems in a global analysis may have a much greater impact on constraining the parameters than would improvement of experimental bounds in systems with current results. 

For example, ThO provides such a tight correlation of $d_e$ and $C_S$, as shown in Fig.~\ref{fig:Paramagnetics}, that narrowing the experimental upper and lower limits without improvements to the other experiments does not significantly improve the bounds on $d_e$ and $C_S$. Adding a degree of freedom, such as a result in Fr, with ${\alpha_{C_S}/\alpha_{d_e}}\approx 1.2\times 10^{-20}$~\cite{rf:Dzuba2011},  could significantly tighten the bounds. Similarly, a result in an octupole-deformed system, {\it e.g.} $^{225}$Ra or $^{221}$Rn/$^{223}$Rn would add a degree of freedom and over-constrain the the set of parameters $C_T$, $\gpbz$, $\gpbo$ and $\bar d_n$. Due to the nuclear structure enhancement of the Schiff moments of such systems, their inclusion in a global analysis could have a substantial impact on the $\gpbi$ as well as on  $C_T$. In contrast , the projected 100-fold improvement in $^{129}$Xe (not octupole-deformed) would have an impact primarily on $C_T$. In the last line of Table~\ref{tab:Outlook}, we optimistically consider the long term prospects with the neutron and $^{129}$Xe improvements and the octupole-deformed systems. The possibility of improvements to TlF, for example with a cooled molecular beam~\cite{rf:Hunter2012} or another molecule will, of course, enhance the prospects.

From a theoretical perspective, it is interesting to consider the theoretical implications of the present and prospective global analysis results. Perhaps, not surprisingly, the resulting constraints on various underlying CPV sources are weaker than under the \lq\lq single-source" assumption. For example, from the limit on $\gpbz$ in Table~\ref{tab:global} and the \lq\lq reasonable range" for the hadronic matrix element computations given in Ref.~\cite{Engel:2013lsa}, we obtain $|{\bar\theta}|\leq{\bar\theta}_\mathrm{max}$, with
\be
2\times 10^{-7} \lsim {\bar\theta}_\mathrm{max} \lsim 1.6\times 10^{-6}\qquad\qquad \mathrm{(global)}
\ee
a constraint considerably weaker than the order $10^{-10}$ upper bound obtained from the neutron or $^{199}$Hg EDM under the \lq\lq single-source" assumption. Similarly, for the dimensionless, isoscalar quark chromo-EDM, the $\gpbz$ bounds imply 
\be
\label{eq:cedm}
{\tilde\delta_q^{(+)}}\, \left(\frac{v}{\Lambda}\right)^2 \lsim 0.01 \ \ \ .
\ee
where we have used the upper end of the hadronic matrix element range given in Ref.~\cite{Engel:2013lsa}. 
Since the quark chromo-EDMs generally arise at one-loop order and may entail strongly interacting virtual particles, we may translate the range in Eq.~(\ref{eq:cedm}) into a range on the BSM mass scale $\Lambda$ by taking ${\tilde\delta}_q^{(+)}\sim \sin\phi_\mathrm{CPV} \times( \alpha_s/4\pi)$ where $\phi_\mathrm{CPV}$ is a CPV phase to obtain
\be
\Lambda\gsim  (2\ \mathrm{TeV})\times \sqrt{\sin\phi_\mathrm{CPV}} \qquad\qquad \mathrm{Isoscalar \ quark\ chromo-EDM\ (global)}\ \ \ .
\ee
We note, however that given the considerable uncertainty in the hadronic matrix element computation these bounds may be considerably weaker\footnote{The uncertainty for the quark CEDM is substantially larger than for those pertaining to ${\bar\theta}$ owing, in the latter case, to the constraints from chiral symmetry as discussed in Ref.~\cite{Engel:2013lsa}.}.  

For the paramagnetic systems, the present mass reach may be substantially greater. For the electron EDM, we again make the one-loop assumption for illustrative purposes, taking $\delta_e \sim \sin\phi_\mathrm{CPV} \times (\alpha/4\pi)$ so that
\be
\Lambda\gsim (1.5\ \mathrm{TeV}) \times \sqrt{\sin\phi_\mathrm{CPV}}  \qquad\qquad\mathrm{Electron\ EDM\ (global)}
\ee
The scalar (quark) $\times$ pseudscalar (electron) interaction leading to a non-vanishing $C_S$ may arise at tree-level, possibly generated by exchange of a scalar particle that does not contribute to the elementary fermion mass through spontaneous symmetry-breaking. In this case, taking
$\mathrm{Im}\, C_{eq}^{(-)}\sim 1$ and using the bound in Table~\ref{tab:global} gives
\be
\Lambda\gsim (1300\ \mathrm{TeV})\times \sqrt{\sin\phi_\mathrm{CPV}} \qquad\qquad C_S\ \mathrm{(global)}
\ee
Under the \lq\lq single-source" assumption, these lower bounds become even more stringent. 

Due to the quadratic dependence of the CPV sources on $(v/\Lambda)$, an order of magnitude increase in sensitivity to any of the hadronic parameters will extend the mass reach by roughly a factor of three. In this respect, achieving the prospective sensitivities for new systems such as Fr and combinations of diamagnetic systems such including the neutron,  $^{129}$Xe and octupole-deformed systems as indicated in Table~\ref{tab:Outlook} would lead to significantly greater mass reach. Achieving these gains, together with the refinements in nuclear and hadronic physics computations needed to translate them into robust probes of underlying CPV sources, 
lays out the future of EDM research in probing BSM Physics.

\begin{table}
\centering
\begin{tabular}{||c|c|c|c|c|c|c|c|c||}
\hline\hline
\multicolumn{3}{||c|}{ }  &  &  &  &  &  &   \\
\multicolumn{3}{||c|}{ }  & $d_e$ (e-cm) & $C_S$ & $C_T$ &$\gpbz$ & $\gpbo$& $\bar d_n$ (e-cm) \\
\hline
\multicolumn{3}{||c|}{Current Limits (95\%)} & $5.4\times 10^{-27}$& $4.5\times 10^{-7}$ & $2\times 10^{-6}$ & $8\times 10^{-9}$ & $1.2\times 10^{-9}$& $12\times 10^{-23}$ \\
\hline\hline
System & Current (e-cm) & Projected   & \multicolumn{6}{c||}{Projected sensitivity}  \\
\hline
ThO   & $5\times 10^{-29}$ &  $5\times 10^{-30}$ &$4.0\times 10^{-27}$& $3.2\times 10^{-7}$  & $ $ &$ $ & $ $& $ $\\
\hline
Fr  & &  $d_e<10^{-28}$ & $2.4\times 10^{-27}$ & $1.8\times 10^{-7}$ & $ $ &$ $ & $ $& $ $\\
\hline
$^{129}$Xe & $3\times 10^{-27}$ & $3\times 10^{-29}$ & & &  $3\times 10^{-7}$ &$3\times 10^{-9} $ & $1\times 10^{-9} $& $5\times 10^{-23}$\\
\hline
Neutron/Xe & $2\times 10^{-26}$ & $10^{-28}$/$3\times 10^{-29}$ & & &  $1\times 10^{-7}$  &$1\times 10^{-9}$ & $4\times 10^{-10} $& $2\times 10^{-23}$\\
\hline
Ra & & $10^{-25}$ & &  & $5\times 10^{-8}$ & $4\times 10^{-9}$&  $1\times 10^{-9} $ & $6\times 10^{-23}$ \\
\hline
" & & $10^{-26}$ & &  & $1\times 10^{-8}$ & $1\times 10^{-9}$&  $3\times 10^{-10} $ & $2\times 10^{-24}$ \\
\hline
Neutron/Xe/Ra &  & $10^{-28}$/$3\times 10^{-29}$/$10^{-27}$ & & &  $6\times 10^{-9}$  &$9\times 10^{-10}$ & $3\times 10^{-10} $& $1\times 10^{-24}$\\
\hline\hline
\end{tabular}
\caption{\label{tab:Outlook} Anticipated limits (95\%) on P-odd/T-odd physics contributions for scenarios for improved experimental precision compared to the current  limits listed in the first line using best values for coefficients in Table~\ref{tb:paramagnetics} and \ref{tb:diamagnetics}. We assume $\alpha_{g_\pi^1}$ for $^{199}$Hg is $1.6\times 10^{-17}$. For the octupole deformed systems ($^{225}$Ra and $^{221}$Rn/$^{223}$Rn) we specify the contribution of $^{225}$Ra. The  Schiff moment for Rn isotopes may be an order of magnitude smaller than for Ra, so for Rn one would require 10$^{-26}$ and 10$^{-27}$ for the fifth and sixth lines to achieve comparable sensitivity to that listed for Ra.}\end{table}









\vskip 0.25in

\noindent\textbf{Acknowledgements} The authors thank P. Fierlinger and G. Gabrielse for useful discussions and both the Excellence Cluster Universe at the Technical University Munich and the Kavli Institute for Theoretical Physics where a portion of this work was completed. MJRM was supported in part by U.S. Department of Energy contracts DE-SC0011095 and DE-FG02-08ER4153,  by the National Science Foundation under Grant No. NSF PHY11-25915, and by the Wisconsin Alumni Research Foundation. TEC was supported in part by the U.S. Department of Energy grant DE FG02 04 ER41331.


\bibliographystyle{h-physrev3.bst}
\bibliography{EDMGlobalRefs}



\end{document}